\documentclass[%
 reprint,
 amsmath,amssymb,
 aps,
]{revtex4-2}
\pdfoutput=1
\usepackage{amsmath}
\usepackage{graphicx}
\usepackage{dcolumn}
\usepackage{bm}
\usepackage{physics}
\usepackage{tikz-feynman}
\tikzfeynmanset{
every blob={/tikz/fill=white!30,/tikz/inner sep=2pt},
}

\begin{document}

\preprint{APS/123-QED}

\title{Limits on the diffractive mass in strong coherent $\gamma^*$-nucleus scattering}

\author{A.\@ H.\@ Mueller} \email{ahm4@columbia.edu}
\affiliation{%
 Department of Physics \\
 Columbia University \\
 New York, NY 10027
}%

\date{\today}

\begin{abstract}
An evolution equation for diffractive production with a definite rapidity gap is given. Coherent $\gamma^*A$ collisions in the unitarity (saturation) region are studied with the conclusion that the diffractive mass is always on the order of the $\gamma^*$ virtuality, $Q$, when the scattering is strong. Deep in the saturation region diffractive masses significantly greater than $Q$ are strongly suppressed by a Levin-Tuchin mechanism.
\end{abstract}

\maketitle


\section{Introduction} \label{sec:1}

The focus of this note is coherent diffraction in $\gamma^*$-nucleus reactions. By coherent we mean that the struck nucleus remains in its ground state after the collision, and by diffraction we suppose there is a rapidity gap between the nucleus and the produced hadronic system. (See appendix \ref{app:A} for a brief review of how we are using rapidities.)

In the projectile frame, where the nucleus is at rest and the momentum of the $\gamma^*$ is $q = (q_+, q_- = -Q^2/2q_+, \underline{0})$ with $q_+$ large, the $\gamma^*$ emits a quark-antiquark ($q\bar{q}$) pair which then evolves and scatters on the target nucleus. If the transverse momentum of the quark (or antiquark) is $p_\perp$ at the time of emission from the $\gamma^*$, and if $p_\perp < Q_S(y)$ with $Q_S$ the nuclear saturation momentum at the rapidity of the scattering, the scattering is said to be in the saturation region where one expects strong scattering. If the quark and antiquark are in the aligned jet region \cite{cite1, cite2} one can have $p_\perp < Q_S$ with $Q \gg Q_S$ and it is this region that we shall focus on.

The main result to be developed in this note is that deep in the saturation region the evolution of the $q\bar{q}$ system is such that the diffractive mass is never significantly larger than $Q$, the virtuality of the $\gamma^*$. That is, large mass ($M > Q$) diffraction is small in inelastic electron scattering when the $q\bar{q}$ system coming from the $\gamma^*$ is in the saturation region giving strong scattering. In an old-fashioned language, the triple pomeron coupling \cite{cite3, cite4, cite5} is zero in QCD.

In Sec.~\ref{sec:2} we review the Kovchegov-Levin \cite{cite6, cite7} equation for diffraction in QCD, and in Sec.~\ref{sec:3} we develop a slightly different evolution equation where the gap size is fixed during the evolution. We then show that if we start evolving the $q\bar{q}$ dipole coming from the $\gamma^*$) (the $x_{01}$ dipole) from the rapidity $y_0$ (the gap size) the diffractive cross section and the diffractive mass both increase as the rapidity, $y$, of the dipole increases beyond $y_0$. This increase continues until the dipole reaches $\bar{y}$ the saturation momentum for the dipole scattering on the nucleus. As one increases the dipole ($x_{01}$) rapidity beyond $\bar{y}$ there is a strong decrease in the diffraction cross section which decrease follows a Levin-Tuchin formula \cite{cite7, cite8, cite9}. Thus when the rapidity is very large the diffractive cross section becomes arbitrarily small when the gap size $y_0$ is less than $\bar{y}$.

In Sec.~\ref{sec:4} we allow $y_0$ to also be greater than $\bar{y}$ and then find for large $y$, using an equation \eqref{eq:12} complimentary to \eqref{eq:4}, that it is not possible to emit a gluon at $y_0$ to fix the rapidity gap to be $y_0$, so again diffraction is very small. Only when $y_0 = y$, that is the diffractive mass is equal to $Q$, is it possible to have strong diffraction where the produced particles are either the $q\bar{q}$ jets coming from the $\gamma^*$ or, in the aligned jet configuration, a DGLAP evolution of these particles. In any case the diffractive mass is on the order of $Q$ and the diffractive particles all have rapidity near $y$ with the total coherent cross section equal to the inelastic cross section.

Finally we give a simple, but heuristic, argument that in a strong scattering situation in $\gamma^*$-nucleus scattering the mass of the diffractive system is on the order of $Q$. In the projectile frame the time scale for the $\gamma^*$ to emit a $q\bar{q}$ pair is $\tau_q = 2q_+/Q^2$ with $q_+$ the longitudinal momentum of the $\gamma^*$. The mass of that (virtual) pair is on the order of $Q$. If later emissions are at times much shorter than $\tau_q$ we may view the process in terms of the evolution and scattering of this ($q\bar{q}$) dipole on the nucleus which has $S=0$ if we are in the saturation region. This $S=0$ means that the elastic (diffractive) process is just two jet production where the two jets have a combined mass on the order of $Q$. If the ``later'' emissions occur over a time $\tau_q$ (as the original $\gamma^*\to q\bar{q}$ does), and this is the situation for DGLAP evolution of the original $q\bar{q}$ pair, the the process can be viewed as the scattering of the DGLAP system on the nucleus which also will have $S=0$ giving an ``elastic'' or diffractive production of the whole DGLAP system. But the DGLAP system of $q\bar{q}$ + gluons also has a mass on the order of $Q$ and so the diffractive mass is again $Q$.

\section{Review of Kovchegov-Levin (KL) equation} \label{sec:2}

In 2000 Y.\@ Kovchegov and E.\@ Levin \cite{cite6, cite7} wrote a general equation for diffractive dissociation in QCD using the dipole picture of high energy scattering. We shall briefly review that equation using the notation in the book by Kovchegov and Levin \cite{cite7}. $N^D(x_{01},y,y_0)$ is the diffractive cross section per unit area for the scattering of a dipole of transverse size $x_{01} = \abs{\underline{x}_0 - \underline{x}_1}$ and having rapidity $y$ on, say, a large nucleus at rest. The gap is greater than or equal to $y_0$. In this (coherent) diffractive scattering the nucleus does not break up and there are no particles produced having rapidities between 0 and $y_0$. Particles produced between $y_0$ and $y$ go toward making up the mass of the diffractively produced system of quarks and gluons. Also defining $N(x_{01},y)$ as the elastic scattering of a dipole of size $x_{01}$ and rapidity $y$ on the nuclear target one has the KL equation
\begin{widetext}
\begin{multline} \label{eq:1}
    \pdv{y} N^D(x_{01},y,y_0) = \frac{\alpha N_C}{2\pi^2} \int \frac{\dd[2]{x_2} x_{01}^2}{x_{12}^2 x_{02}^2} \big\{N^D(x_{02},y,y_0) + N^D(x_{12},y,y_0) - N^D(x_{01},y,y_0) + N^D(x_{02},y,y_0) N^D(x_{12},y,y_0) \\
    - 2 N(x_{02},y) N^D(x_{12},y,y_0) - 2 N(x_{12},y)  N^D(x_{02},y,y_0) + 2 N(x_{12},y) N(x_{02},y)\big\} \,.
\end{multline}
\end{widetext}

The $N^D$ above does not have a definite gap size. The requirement is that events must have a gap size greater than or equal to $y_0$. For our purposes it is important to have an exact specification of the gap size in order to precisely characterize events at high energy. To better appreciate the subtleties in using \eqref{eq:1} consider the very high energy limit, $y$ becoming very large with $y_0$ fixed, of that equation. At very large $y$, $N(x_{12},y)$ and $N(x_{02},y)$ become equal to 1 and \eqref{eq:1} becomes
\begin{multline} \label{eq:2}
    \pdv{y} N^D(x_{01},y,y_0) = \frac{\alpha N_C}{2\pi^2} \int \frac{\dd[2]{x_2} x_{01}^2}{x_{12}^2 x_{02}^2} \\
    \times \big\{2 - N^D(x_{02},y,y_0) - N^D(x_{12},y,y_0) \\
    - N^D(x_{01},y,y_0) + N^D(x_{12},y,y_0) N^D(x_{02},y,y_0) \big\}
\end{multline}
which has a solution
\begin{equation} \label{eq:3}
    N^D(x_{ij},y,y_0) \xrightarrow[y\to\infty]{} 1 \,.
\end{equation}
However, in this solution it happens that the only type of events present are elastic dipole scattering events as we shall see in the next sections. The detailed nature of the event, for example elastic or inelastic projectile scattering is difficult to see from \eqref{eq:1} and \eqref{eq:2}.

\section{An evolution equation for fixed rapidity gap} \label{sec:3}

In this section we are going to give an equation for diffraction, $N^D_t(x_{01},y,y_0)$, where the subscript $t$ means that when $y > y_0$ a gluon in the final state is actually tagged to have a momentum exactly equal to $y_0$.

\subsection{The equation} \label{sec:3.1}

\begin{figure*}[htbp]
\centering
\begin{equation*}
\pdv{y}
\vcenter{\hbox{
\begin{tikzpicture}
\begin{feynman} 
\vertex (tl);
\vertex [right=1cm of tl] (tm);
\vertex [right=1cm of tm] (tr);
\vertex [below=1cm of tl] (bl);
\vertex [right=1cm of bl] (bm);
\vertex [right=1cm of bm] (br);
\vertex [above=0.5cm of tm] (tt);
\vertex [below=0.5cm of bm] (bb);

\diagram* {
  (tl) -- (tm) -- [edge label=\(\underline{x}_1\), near end] (tr),
  (bl) -- (bm) -- [edge label'=\(\underline{x}_0\), near end] (br),
  (tm) -- [half right] (bm) -- [half right] (tm),
};

\diagram* {
  (tt) -- (bb),
};

\vertex [below=0.23cm of tm] {\(N^D_t\)};
\end{feynman}
\end{tikzpicture}
}}
=
\vcenter{\hbox{
\begin{tikzpicture}
\begin{feynman} 
\vertex (tl);
\vertex [right=1cm of tl] (tm);
\vertex [right=1cm of tm] (tr);
\vertex [below=1.5cm of tl] (bl);
\vertex [right=1cm of bl] (bm);
\vertex [right=1cm of bm] (br);
\vertex [below=0.75cm of tl] (cl);
\vertex [right=1cm of cl] (cm);
\vertex [right=1cm of cm] (cr);
\vertex [above=0.5cm of tm] (tt);
\vertex [below=0.5cm of bm] (bb);

\diagram* {
  (tl) -- (tm) -- [edge label=\(\underline{x}_1\), near end] (tr),
  (bl) -- (bm) -- [edge label'=\(\underline{x}_0\), near end] (br),
  (cl) -- [photon] (cm) -- [photon] (cr),
  (tm) -- [half right] (cm) -- [half right] (tm),
  (cm) -- [half right] (bm) -- [half right] (cm),
};

\diagram* {
  (tt) -- (bb),
};

\vertex [below=0.1cm of tm] {\(N^D_t\)};
\vertex [below=0.1cm of cm] {\(S^D\)};
\vertex [right=0.05cm of cr] {\(\underline{x}_2\)};
\end{feynman}
\end{tikzpicture}
}}
+
\vcenter{\hbox{
\begin{tikzpicture}
\begin{feynman} 
\vertex (tl);
\vertex [right=1cm of tl] (tm);
\vertex [right=1cm of tm] (tr);
\vertex [below=1.5cm of tl] (bl);
\vertex [right=1cm of bl] (bm);
\vertex [right=1cm of bm] (br);
\vertex [below=0.75cm of tl] (cl);
\vertex [right=1cm of cl] (cm);
\vertex [right=1cm of cm] (cr);
\vertex [above=0.5cm of tm] (tt);
\vertex [below=0.5cm of bm] (bb);

\diagram* {
  (tl) -- (tm) -- [edge label=\(\underline{x}_1\), near end] (tr),
  (bl) -- (bm) -- [edge label'=\(\underline{x}_0\), near end] (br),
  (cl) -- [photon] (cm) -- [photon] (cr),
  (tm) -- [half right] (cm) -- [half right] (tm),
  (cm) -- [half right] (bm) -- [half right] (cm),
};

\diagram* {
  (tt) -- (bb),
};

\vertex [below=0.1cm of tm] {\(S^D\)};
\vertex [below=0.1cm of cm] {\(N^D_t\)};
\vertex [right=0.05cm of cr] {\(\underline{x}_2\)};
\end{feynman}
\end{tikzpicture}
}}
+
\vcenter{\hbox{
\begin{tikzpicture}
\begin{feynman} 
\vertex (tl);
\vertex [right=1.5cm of tl] (tm);
\vertex [right=1cm of tm] (tr);
\vertex [below=1cm of tl] (bl);
\vertex [right=1.5cm of bl] (bm);
\vertex [right=1cm of bm] (br);
\vertex [below=0.5cm of tl] (cl);
\vertex [right=0.85 cm of cl] (cm);
\vertex [above=0.5cm of tm] (tt);
\vertex [below=0.5cm of bm] (bb);

\diagram* {
  (tl) -- [edge label=\(\underline{x}_1\), near start] (tm) -- (tr),
  (bl) -- [edge label'=\(\underline{x}_0\), near start] (bm) -- (br),
  (cl) -- [photon] (cm),
  (tm) -- [half right] (bm) -- [half right] (tm),
};

\diagram* {
  (tt) -- (bb),
};

\vertex [below=0.23cm of tm] {\(N^D_t\)};
\vertex [left=0.05cm of cl] {\(\underline{x}_2\)};
\end{feynman}
\end{tikzpicture}
}}
\end{equation*}
\caption{ \label{fig:1}}
\end{figure*}

The equation equivalent and complimentary to \eqref{eq:1} is
\begin{widetext}
\begin{equation} \label{eq:4}
    \pdv{y} N^D_t(x_{01},y,y_0) = \frac{\alpha N_C}{2\pi^2} \int \frac{\dd[2]{x_2} x_{01}^2}{x_{12}^2 x_{02}^2} \big\{N^D_t(x_{12},y,y_0) S^D(x_{02},y,y_0) + N^D_t(x_{02},y,y_0) S^D(x_{12},y,y_0) - N^D_t(x_{01},y,y_0) \big\}
\end{equation}
\end{widetext}
where
\begin{equation} \label{eq:5}
    S^D(x_{01},y,y_0) = 1 - 2N(x_{01},y) + N^D(x_{01},y,y_0)
\end{equation}
as in KL \cite{cite6, cite7} and $N(x_{01},y)$, $N^D(x_{01},y,y_0)$ are as previously used in \eqref{eq:1}. The initial condition for \eqref{eq:4} is the same as for \eqref{eq:1}. In \eqref{eq:4} $N^D_t(x_{01},y,y_0)$ corresponds to a rapidity gap exactly equal to $y_0$. The gluons between $y_0$ and $y$ go toward creating a diffractive mass $M$ having size
\begin{equation*}
    M^2 = Q^2 e^{y-y_0} \,.
\end{equation*}

It is easy to check the validity of \eqref{eq:4} by inspection. By increasing the rapidity of the dipole $x_{01}$ two dipoles $x_{02}$ and $x_{12}$ are created as usual. In the first term on the right hand side of \eqref{eq:4} dipole $x_{12}$ scatters diffractively on the target with a minimum energy gluon at $y_0$ while dipole $x_{02}$ may (i) not scatter at all, (ii) scatter elastically in the amplitude or complex conjugate amplitude, or (iii) diffractively scatter with a diffractive gap greater or equal to $y_0$. These three terms correspond to the three terms in \eqref{eq:5} for $S^D$. The second term on the right hand side of \eqref{eq:4} is the same as the first term with the roles of $x_{02}$ and $x_{12}$ exchanged. The final term on the right hand side of \eqref{eq:4} is the usual probability conserving term. It is straightforward to derive \eqref{eq:4} from \eqref{eq:1} by noting that
\begin{equation} \label{eq:6}
    N^D_t(x_{01},y,y_0) = -\pdv{y_0} N^D(x_{01},y,y_0) \,,
\end{equation}
in which case the $\pdv*{y_0}$ operator on \eqref{eq:1} immediately gives \eqref{eq:4}. Eq.~\eqref{eq:4} is illustrated in Fig.~\ref{fig:1}.

\subsection{The weak and strong scattering limits of \eqref{eq:4}} \label{sec:3.2}

\begin{figure}[htbp]
\centering
\begin{tikzpicture}
\begin{feynman}
\vertex (o);
\vertex [right=4cm of o] (xmark);
\vertex [right=5cm of o] (x1);
\vertex [right=1cm of x1] (x2);
\vertex [above=2cm of o] (y0mark);
\vertex [right=6cm of y0mark] (y0end);
\vertex [above=2cm of y0mark] (ymark);
\vertex [left=0.05cm of ymark] (yl);
\vertex [right=0.05cm of ymark] (yr);
\vertex [above=1cm of ymark] (y1);
\vertex [above=1cm of y1] (y2);
\vertex [right=1cm of o] (l1);
\vertex [right=2.33333cm of y0mark] (l2);
\vertex [right=3.66667cm of ymark] (l3);
\vertex [left=0.05cm of l3] (ll);
\vertex [right=0.05cm of l3] (lr);
\vertex [right=4.25cm of y1, dot] (l4) {};
\vertex [left=2cm of l4] (h1);
\vertex [right=5cm of y2] (l5);
\vertex [below=3cm of l4, dot] (v1) {};
\vertex [below=1cm of l4, dot] (v2) {};
\vertex [above=2cm of l4, dot] (v3) {};

\diagram* {
  (o) -- (x1) -- [momentum'] (x2),
  (o) -- (y1) -- [momentum] (y2),
  (y0mark) -- (y0end),
  (l1) -- (l5),
  (v1) -- (v2) -- (l4) -- (v3),
  (h1) -- [scalar] (l4),
  (yl) -- (yr),
  (ll) -- (lr),
};

\vertex [below=5.4cm of l4] {\(\ln(\frac{1}{x_\perp^2})\)};
\vertex [left=0cm of y0mark] {\(y_0\)};
\vertex [left=0.1cm of y1] {\(y\)};
\vertex [below=0.3cm of v1] {\(1\)};
\vertex [right=0.3cm of v2] {\(2\)};
\vertex [right=0.3cm of l4] {\(3\)};
\vertex [right=0.3cm of v3] {\(y\)};
\vertex [above right=0.05cm of l5] {\(\ln Q_S^2(y)\)};
\vertex [left=0.05cm of h1] {\(\bar{y}\)};

\end{feynman}
\end{tikzpicture}
\caption{ \label{fig:2}}
\end{figure}

Consider the picture in Fig.~\ref{fig:2} illustrating the different domains of high energy diffractive scattering of a dipole $x_{01}$ on a large nucleus. Imagine starting by scattering the dipole $x_{01}$ elastically on the nucleus at point 1 in the figure where the dipole has rapidity $y_0$. The gap is $y_0$ in the reaction. Now increase the rapidity of the dipole from point 1 vertically to point 2 keeping a softest gluon at $y_0$ to insure the rapidity gap remain exactly $y_0$. The scattering is always coherent leaving the nucleus in its ground state. The diagonal line in Fig.~\ref{fig:1} represents $\ln Q_S^2(y)$, the nuclear saturation momentum at rapidity $y$. Thus so long as the dipole $x_{01}$ is to the right of the $\ln Q_S^2$ line the scattering of the dipole on the nucleus is weak. When the dipole scattering is weak $S^D \simeq 1$ and \eqref{eq:4} just becomes the dipole version of BFKL evolution, during which evolution the diffractive mass is increasing, with the rapidity gap fixed at $y_0$. Thus the evolution from 1 to 2 in Fig.~\ref{fig:2} is just BFKL evolution increasing the amplitude with the diffractive mass.

As one increases the rapidity of the dipole beyond point 2 a strong change in the evolution given by \eqref{eq:4} occurs when one reaches the nuclear saturation line at point 3. At, and beyond point 3, the quantity $S^D(x_{01},y_3,y_0) \simeq 0$ and \eqref{eq:4} becomes
\begin{equation} \label{eq:7}
    \pdv{y} N^D_t(x_{01},y,y_0) \simeq -\frac{\alpha N_C}{2\pi^2} \int \frac{\dd[2]{x_2} x_{01}^2}{x_{12}^2 x_{02}^2} N^D_t(x_{01},y,y_0)
\end{equation}
so long as $x_{12}$ and $x_{02}$ do not become too small. As we move above point 3 in Fig.~\ref{fig:2} the dominant contribution comes from $x_{12}$ (or $x_{02}$) becoming smaller than $x_{01}$. Taking $x_{12}/x_{01} \ll 1$ and including a factor of 2 for $x_{02}/x_{01} \ll 1$ the more precise form of \eqref{eq:7} is
\begin{equation} \label{eq:8}
\begin{split}    
    \pdv{y} N^D_t(x_{01},y,y_0) &= -\frac{\alpha N_C}{2\pi^2} 2\pi \int_{1/Q_S^2(y)}^{x_{01}^2} \frac{\dd{x_{12}^2}}{x_{12}^2} N^D_t(x_{01},y,y_0) \\
    &= -\frac{\alpha N_C}{\pi} \ln\qty[Q_S^2(y) x_{01}^2] N^D_t(x_{01},y,y_0) \,.
\end{split}
\end{equation}
Calling $\bar{y}$ the value of $y$ at point 3, and integrating \eqref{eq:8} between $\bar{y}$ and $y$ gives
\begin{equation} \label{eq:9}
    N^D_t(x_{01},y,y_0) = N^D_t(x_{01},\bar{y},y_0) e^{-\qty(\frac{\alpha N_C}{\pi})^2 \frac{\chi(\lambda_0)}{1-\lambda_0} (y-\bar{y})^2}
\end{equation}
where we have used
\begin{equation} \label{eq:10}
    Q_S^2(y) = Q_S^2(\bar{y}) e^{ \frac{2\chi(\lambda_0)}{1-\lambda_0} \frac{\alpha N_C}{\pi} (y-\bar{y})}
\end{equation}
with $\chi(\lambda)$ the BFKL characteristic function and $\frac{\chi'(\lambda_0)}{\chi(\lambda_0)} = -\frac{1}{1-\lambda_0}$. \eqref{eq:8} and \eqref{eq:9} are the forms of a Levin-Tuchin equation and its solution \cite{cite8, cite9, cite10}. The Levin-Tuchin equation occurs when real emissions are forbidden in small-$x$ evolution and the resulting suppression (the exponential decrease in $y$ in \eqref{eq:9}) occurs due to virtual terms in the evolution, as in \eqref{eq:7} above.

\section{Limits on the diffractive mass in the black disc limit} \label{sec:4}

\subsection{The process} \label{sec:4.1}

Now we are going to systematically examine the contstraints on the diffractive mass when scattering is in the black disc regime. Equations \eqref{eq:7}-\eqref{eq:10} are a start in that direction. To make the discussion more physical we suppose that the process being considered is coherent $\gamma^*$ scattering on a large nucleus at high energy. For $\gamma^*$ having transverse polarization the coherent diffractive cross section is given as
\begin{widetext}
\begin{equation} \label{eq:11}
    \frac{M^2 \dd{\sigma^{\gamma^*A}}}{\dd[2]{b} \dd{M^2}} = 2 N_C \alpha_{em} e_f^2 \int \frac{\dd[2]{x}}{(2\pi)^2} \int_0^1 \dd{z} \qty[z^2 + (1-z)^2] \bar{Q}^2 K_1^2(\bar{Q}x_\perp) N^D_t(x_\perp,y,y_0)
\end{equation}
\end{widetext}
for diffraction initated by $\gamma^* \to q_f + \bar{q}_f$. The nontrivial part of \eqref{eq:11} is given by $N^D_t(x_\perp,y,y_0)$. In the last section we moved $y$ from $y_0$ along the vertical line in Fig.~\ref{fig:2} with $y_0$ always below the saturation line $\ln Q_S^2(y)$ while we found BFKL evolution of $N_t^D$ when $y$ was below the saturation line and a Levin-Tuchin type of result \eqref{eq:9} when $y$ went above the saturation line, at $\bar{y}$ in the region where the size, $x_\perp$, of the $q\bar{q}$ dipole coming from the $\gamma^*$ was firmly in the black disc regime. Now we also are going to vary $y_0$ from $y_0 < \bar{y}$ (below the saturation line) to $y_0 > \bar{y}$ (above the saturation line). We begin with \eqref{eq:8} and \eqref{eq:9} expressing the result when $y_0 < \bar{y} < y$.

\subsection{Diffraction with $y_0 < \bar{y} < y$} \label{sec:4.2}

When $y$ is significantly above $\bar{y}$, that is when $\alpha(y-\bar{y}$ is large, one is deep in the saturation regime for the scattering and \eqref{eq:9} gives a very small diffractive cross section with the suppression factor independent of $y_0$. The diffractive events are somewhat unusual starting from $y$ and looking at decreasing rapidities one finds $q$ and $\bar{q}$ jets at $y$. Then there is a rapidity gap between $y$ and $\bar{y}$. At even lower rapidities there are soft gluons, in a BFKL distribution, between $\bar{y}$ and $y_0$ and, of course, the rapidity gap below $y_0$ which was our original imposition on the events to call them diffractive.

The first rapidity gap, between $\bar{y}$ and $y$, which we did not impose follows from \eqref{eq:7} which came from \eqref{eq:6} using $S^D = 0$ when $y > \bar{y}$. There is a simple argument as to why this had to happen. Consider the first term on the right hand side of \eqref{eq:4} in which the gluon at $y_0$ is contained in $N^D_t(x_{12},y_1,y_0)$. Then the $x_{02}$ dipole passing through the nucleus will surely break up the nucleus since its rapidity is so large. This means that it is not possible to emit the gluon at $\underline{x}_2$ with rapidity above $\bar{y}$ and not break up the nucleus. This gives a very small probability of a gap below $y_0$. Thus if we require that there be a soft gluon at $y_0$ giving the size of the rapidity gap defining the diffractive events, and if $y$ is significantly greater than $\bar{y}$ then all diffraction is strongly suppressed.

\subsection{Trying to take $y_0 > \bar{y}$} \label{sec:4.3}

\begin{figure*}[htbp]
\centering
\begin{equation*}
\vcenter{\hbox{
\begin{tikzpicture}
\begin{feynman} 
\vertex [particle] (p) {\(\gamma^*\)};
\vertex [right=of p] (l);
\vertex [below right=1cm of l] (b1);
\vertex [right=1.5cm of b1] (b2);
\vertex [right=1.0cm of b2] (b3);
\vertex [right=0.5cm of b3] (b4);
\vertex [above right=1cm of l] (t1);
\vertex [right=1.5cm of t1] (t2);
\vertex [right=1.0cm of t2] (t3);
\vertex [right=0.5cm of t3] (t4);
\vertex [right=0.5cm of l] (m1);
\vertex [right=2.0cm of m1] (m2);
\vertex [right=1.0cm of m2] (m3);
\vertex [right=0.5cm of m3] (m4);
\vertex [below=0.5cm of b2, blob] (n) {\(A\)};
\vertex [above=0.4cm of b2] (nt);
\vertex [below left=0.7cm of n] (nl);
\vertex [below right=0.7cm of n] (nr);

\diagram* {
  (l) -- [quarter right] (b1) -- (b2) -- (b3) -- [fermion, edge label'=\(\underline{x}_1\), near start] (b4),
  (l) -- [quarter left] (t1) -- (t2) -- (t3) -- [anti fermion, edge label=\(\underline{x}_0\), near start] (t4),
  (m1) -- [gluon] (m3),
  (p) -- [photon] (l),
  (nt) -- [gluon] (n),
  (nl) -- [double] (n),
  (nr) -- [double] (n),
};

\vertex [right=0.05cm of m3] {\(\underline{x}_2\)};

\node[above=-0.1cm of current bounding box.north] {(a)};
\end{feynman}
\end{tikzpicture}
}}
+
\vcenter{\hbox{
\begin{tikzpicture}
\begin{feynman} 
\vertex [particle] (p) {\(\gamma^*\)};
\vertex [right=of p] (l);
\vertex [below right=1cm of l] (b1);
\vertex [right=1.5cm of b1] (b2);
\vertex [right=1.0cm of b2] (b3);
\vertex [right=0.5cm of b3] (b4);
\vertex [above right=1cm of l] (t1);
\vertex [right=1.5cm of t1] (t2);
\vertex [right=1.0cm of t2] (t3);
\vertex [right=0.5cm of t3] (t4);
\vertex [right=0.5cm of l] (m1);
\vertex [right=2.0cm of m1] (m2);
\vertex [right=1.0cm of m2] (m3);
\vertex [right=0.5cm of m3] (m4);
\vertex [below=0.5cm of b2, blob] (n) {\(A\)};
\vertex [above=1.2cm of b2] (nt);
\vertex [below left=0.7cm of n] (nl);
\vertex [below right=0.7cm of n] (nr);

\diagram* {
  (l) -- [quarter right] (b1) -- (b2) -- (b3) -- [fermion, edge label'=\(\underline{x}_1\), near start] (b4),
  (l) -- [quarter left] (t1) -- (t2) -- (t3) -- [anti fermion, edge label=\(\underline{x}_0\), near start] (t4),
  (m1) -- [gluon] (m3),
  (p) -- [photon] (l),
  (nt) -- [gluon] (n),
  (nl) -- [double] (n),
  (nr) -- [double] (n),
};

\vertex [right=0.05cm of m3] {\(\underline{x}_2\)};

\node[above=-0.1cm of current bounding box.north] {(b)};
\end{feynman}
\end{tikzpicture}
}}
\end{equation*}
\begin{equation*}
+
\vcenter{\hbox{
\begin{tikzpicture}
\begin{feynman} 
\vertex [particle] (p) {\(\gamma^*\)};
\vertex [right=of p] (l);
\vertex [below right=1cm of l] (b1);
\vertex [right=0.5 of b1] (b2);
\vertex [right=1.5cm of b2] (b3);
\vertex [right=0.5cm of b3] (b4);
\vertex [right=0.5cm of b4] (b5);
\vertex [above right=1cm of l] (t1);
\vertex [right=1.5cm of t1] (t2);
\vertex [right=1.0cm of t2] (t3);
\vertex [right=0.5cm of t3] (t4);
\vertex [right=0.5cm of l] (m1);
\vertex [right=2.0cm of m1] (m2);
\vertex [right=1.0cm of m2] (m3);
\vertex [right=0.5cm of m3] (m4);
\vertex [below=0.5cm of b2, blob] (n1) {\(A\)};
\vertex [above=1.2cm of b2] (n1t);
\vertex [below left=0.7cm of n1] (n1l);
\vertex [below right=0.7cm of n1] (n1r);
\vertex [below=0.5cm of b3, blob] (n2) {\(A\)};
\vertex [above=0.25cm of b3] (n2t);
\vertex [below left=0.7cm of n2] (n2l);
\vertex [below right=0.7cm of n2] (n2r);

\diagram* {
  (l) -- [quarter right] (b1) -- (b2) -- (b4) -- [fermion, edge label'=\(\underline{x}_1\), near start] (b5),
  (l) -- [quarter left] (t1) -- (t2) -- (t3) -- [anti fermion, edge label=\(\underline{x}_0\), near start] (t4),
  (m1) -- [gluon] (m3),
  (p) -- [photon] (l),
  (n1t) -- [gluon] (n1),
  (n1l) -- [double] (n1),
  (n1r) -- [double] (n1),
  (n2t) -- [gluon] (n2),
  (n2l) -- [double] (n2),
  (n2r) -- [double] (n2),
};

\vertex [right=0.05cm of m3] {\(\underline{x}_2\)};

\node[above=-0.1cm of current bounding box.north] {(c)};
\end{feynman}
\end{tikzpicture}
}}
+
\vcenter{\hbox{
\begin{tikzpicture}
\begin{feynman} 
\vertex [particle] (p) {\(\gamma^*\)};
\vertex [right=of p] (l);
\vertex [below right=1cm of l] (b1);
\vertex [right=0.75cm of b1] (b2);
\vertex [right=1.75cm of b2] (b3);
\vertex [right=0.5cm of b3] (b4);
\vertex [above right=1cm of l] (t1);
\vertex [right=1.5cm of t1] (t2);
\vertex [right=1.0cm of t2] (t3);
\vertex [right=0.5cm of t3] (t4);
\vertex [right=0.5cm of l] (m1);
\vertex [right=2.0cm of m1] (m2);
\vertex [right=1.0cm of m2] (m3);
\vertex [right=0.5cm of m3] (m4);
\vertex [below=0.5cm of b2, blob] (n) {\(A\)};
\vertex [above=0.7cm of b2] (nt);
\vertex [below left=0.7cm of n] (nl);
\vertex [below right=0.7cm of n] (nr);

\diagram* {
  (l) -- [quarter right] (b1) -- (b2) -- (b3) -- [fermion, edge label'=\(\underline{x}_1\), near start] (b4),
  (l) -- [quarter left] (t1) -- (t2) -- (t3) -- [anti fermion, edge label=\(\underline{x}_0\), near start] (t4),
  (m2) -- [gluon] (m3),
  (p) -- [photon] (l),
  (nt) -- [gluon] (n),
  (nl) -- [double] (n),
  (nr) -- [double] (n),
};

\vertex [right=0.05cm of m3] {\(\underline{x}_2\)};

\node[above=-0.1cm of current bounding box.north] {(d)};
\end{feynman}
\end{tikzpicture}
}}
\end{equation*}
\caption{ \label{fig:3}}
\end{figure*}

Now suppose to tray to take $y_0 > \bar{y}$
, that is the gluon emission at $y_0$ is also in the saturation region. The argument of Sec.~\ref{sec:3.2} giving a small probability for soft gluon emission between $\bar{y}$ and $y$ now immediately applies for soft gluons between $y_0$ and $y$. Thus the gluon at $y_0$ is the hardest gluon emitted. But it is also the softest gluon emitted because of our requirement that there be a gap between the coherent nucleus and $y_0$. Thus the diffractive state is just the forward $q$ and $\bar{q}$ along with the gluon at $y_0 > \bar{y}$. We are now going to show that a soft gluon, $\underline{x}_2$, cannot be produced from a dipole $(\underline{x}_0,\underline{x}_1)$ by elastic scattering. The graphs for the amplitude are illustrated in Fig.~\ref{fig:3}, while starting from the dipole $x_{01}$ the process probability of emitting the gluon and the electric scattering is \footnote{Eq.~\eqref{eq:12} has also been found by E.\@ Iancu (unpublished notes)}
\begin{equation} \label{eq:12}
    \pdv{N_{01}}{y} = \frac{\alpha N_C}{2\pi^2} \int \frac{\dd[2]{x_2} x_{01}^2}{x_{12}^2 x_{02}^2} \qty[N_{12} + N_{02} - N_{12}N_{02} - N_{01}]^2
\end{equation}
where the terms $N_{12}$, $N_{02}$, $N_{12}N_{02}$ and $N_{01}$ correspond to the terms (a), (b), (c) and (d) in Fig.~\ref{fig:3}. Since one is in the saturation region \eqref{eq:12} is equal to zero since all the $N$'s, representing elastic scattering on the nucleus, are equal to one. The $N_{01}$ term is a final state emission while the other terms are initial state emissions. Thus it appears that the cancellation given in \eqref{eq:12} is very different from that leading from \eqref{eq:4} to \eqref{eq:7}. The result \eqref{eq:9} following from \eqref{eq:7} comes about because of a \textit{suppression} of real emissions in the initial state of the amplitude for coherent diffraction as explained at the end of Sec.~\ref{sec:4.2} while the inability to emit a gluon at $y_0$, when $y_0 > \bar{y}$, comes from a \textit{cancellation} of initial and final state emissions.

Of course if $y-y_0$ is not large the gluon emitted at $y_0$ will not be soft and the cancellation given in \eqref{eq:12} does not happen. But in this case the diffractive mass is on the order of $Q$ and there is no distinction between $y$ and $y_0$. In addition the emission of the gluon at $y_0$ is down by a factor of $\alpha$ without setting an independent scale for the gap. The only way to emit a gluon around $y$ from the $q\bar{q}$ pair coming from the $\gamma^*$ without losing a factor of $\alpha$ is to have that gluon be part of DGLAP evolution of the $q\bar{q}$ system. Thus starting from the aligned jet configuration of the $q\bar{q}$ system, with $q$ and $\bar{q}$ having transverse momentum $(p+k)_\perp$, with $p_\perp < Q_S < k_\perp < Q$ the softer quark (or antiquark) may emit a gluon of transverse momentum $k_\perp$ and the $\alpha$ of that emission is compensated by the $\dd{k_\perp^2}/k_\perp^2$ integration. Thus a $q\bar{q}g$ system is created with the transverse momentum of one of the quarks at $p_\perp$ and the other quark and gluon close to $k_\perp$. Indeed if $Q$ is large enough there may be several steps of DGLAP evolution. However, all this DGLAP evolution is happening at rapidity on the order of $y$ and the mass of the diffractive system is still, parametrically on the order of $Q$.

\subsection{In summary} \label{sec:4.4}

The various cases that we have considered all lead to the same result. In coherent diffraction of $\gamma^*$ on a nucleus where the $q$ and $\bar{q}$ emitted by the $\gamma^*$ are deep in the saturation region of the nucleus the diffractive mas is always on the order of $Q$. If the $q$ and $\bar{q}$ coming from the $\gamma^*$ each have $p_\perp \sim Q<Q_S$ then, at leading order in $\alpha$ the diffractive system consists only of the $q$ and $\bar{q}$ ``jets''. If the $Q$ of the $\gamma^*$ obeys $Q \gg Q_S$ and if the $q\bar{q}$ system emitted by the $\gamma^*$ is in the aligned jet region and with a transverse momentum less than $Q_S$ then a DGLAP evolved $q\bar{q}$ system may occur, still at a mass on the order of $Q$. No matter how high the energy the coherent diffractive process never produces masses significantly larger than $Q$ if the emission is in the saturation region. This is all an example of the vanishing of the triple pomeron coupling in case $y_0 > \bar{y}$. If one could diffractively produce a large mass system completely in the saturation region then the large mass would be represented by the pomeron while the production amplitude would have an exchanged pomeron between the coherent nuclear scattering and the large diffractive mass. Such a triple pomeron coupling would naturally violate unitarity limits. This was a severe problem in the early days of trying to understand high energy hadronic scattering. Here the vanishing of the triple pomeron coupling seems to occur very naturally in QCD.

\begin{acknowledgments}
I have benefited from many conversations with Edmond Iancu and Feng Yuan while the work was being done. In particular, Eq.~\eqref{eq:12} in the text was also found by Edmond Iancu. This work was supported in part through a DOE grant DE-SC0011941.
\end{acknowledgments}

\appendix

\renewcommand\thefigure{\thesection.\arabic{figure}}

\section{Rapidity definitions} \label{app:A}

\setcounter{figure}{0}

\begin{figure}[htbp]
\centering
\begin{tikzpicture}
\begin{feynman} 
\vertex (p);
\vertex [right=of p] (l);
\vertex [below right=1cm of l] (b1);
\vertex [right=1.0cm of b1] (b2);
\vertex [right=0.5cm of b2] (b3);
\vertex [right=0.5cm of b3] (b4);
\vertex [right=1.0cm of b4] (b5);
\vertex [above right=1cm of l] (t1);
\vertex [right=1.0cm of t1] (t2);
\vertex [right=0.5cm of t2] (t3);
\vertex [right=0.5cm of t3] (t4);
\vertex [right=1.0cm of t4] (t5);
\vertex [below=1cm of b3, blob] (n) {};
\vertex [below left=1.2cm of n] (nl);
\vertex [below right=1.2cm of n] (nr);

\diagram* {
  (l) -- [quarter right] (b1) -- [momentum'=\(p\)] (b2) -- (b3) -- (b4) -- [fermion, momentum=\(p+\delta\)] (b5),
  (l) -- [quarter left] (t1) -- (t2) -- (t3) -- (t4) -- [anti fermion, momentum=\(q-p\), near start] (t5),
  (p) -- [photon, momentum'=\(q\)] (l),
  (n) -- [gluon, momentum'=\(\delta\)] (b3),
  (nl) -- [double, momentum=\(P\)] (n),
  (n) -- [double, momentum=\(P-\delta\)] (nr),
};
\end{feynman}
\end{tikzpicture}
\caption{ \label{fig:A1}}
\end{figure}

\begin{figure}[htbp]
\centering
\begin{tikzpicture}
\begin{feynman} 
\vertex [blob] (n) {};
\vertex [below right=1.5cm of n] (l);
\vertex [below right=1cm of l] (b1);
\vertex [right=1.0cm of b1] (b2);
\vertex [right=0.5cm of b2] (b3);
\vertex [right=0.5cm of b3] (b4);
\vertex [right=1.0cm of b4] (b5);
\vertex [above right=1cm of l] (t1);
\vertex [right=1.0cm of t1] (t2);
\vertex [right=0.5cm of t2] (t3);
\vertex [right=0.5cm of t3] (t4);
\vertex [right=1.0cm of t4] (t5);
\vertex [left=1.2cm of n] (nl);
\vertex [right=1.2cm of n] (nr);
\vertex [below left=1.2cm of b3, particle] (p) {\(\gamma\)};

\diagram* {
  (l) -- [quarter right] (b1) -- [momentum=\(-p\)] (b2) -- (b3) -- (b4) -- [fermion, momentum=\(q-p\)] (b5),
  (l) -- [quarter left] (t1) -- (t2) -- (t3) -- (t4) -- [anti fermion, momentum=\(p+\delta\), near start] (t5),
  (n) -- [quarter right, gluon, momentum'=\(\delta\)] (l),
  (nl) -- [double, momentum=\(P\)] (n),
  (n) -- [double, momentum=\(P-\delta\)] (nr),
  (p) -- [photon, momentum'=\(q\)] (b3),
};
\end{feynman}
\end{tikzpicture}
\caption{ \label{fig:A2}}
\end{figure}

Although our use of rapidity is reasonably conventional we review the essentials of our use of rapidity in this appendix in the context of a lowest order diffractive production of a $q\bar{q}$ pair in coherent $\gamma^*A$ scattering. Figs.~\label{fig:A1} and \label{fig:A2} illustrate projectile and target pictures of the scattering. We define rapidity in terms of the longitudinal (the minus component) momentum in the target frame. Thus the rapidity of the $p+\delta$ line is
\begin{equation}
    y_{p+\delta} = \ln\frac{P_-}{(p+\delta)_-} \simeq \ln\frac{1}{x} \equiv y
\end{equation}
where $2p\cdot q/Q^2 = 1/x$. Similarly
\begin{equation}
    y_{q-p} = \ln\frac{1}{x} + \ln\frac{q_+}{p_+} \simeq \ln\frac{1}{x} + \ln\frac{Q^2}{p_\perp^2}
\end{equation}
while
\begin{equation}
    y_{-p} = \ln\frac{P_-}{-p_-} = \ln\frac{1}{x} \,.
\end{equation}
It is $y_{p+\delta} = \ln 1/x$ that we have called $y$.

In the projectile picture the inverse minus components of the momenta are lifetimes where $\tau_{p+\delta} = 2p_+/p_\perp^2$, $\tau_q=2q_+/Q^2$ etc. Then
\begin{align}
    y_{p+\delta} &= \ln[P_- \tau_{p+\delta}] \\
    y_{q-p} &= \ln[P_- \tau_{q-p}] \\
    y_{-p} &= \ln[P_- \tau_{p}]
\end{align}
The mass of the $q\bar{q}$ pair is given as
\begin{equation}
\begin{split}
    M^2 &= 2(q-p)\cdot(p+\delta) \\
    &\simeq 2(q-p)_+(p+\delta)_- \\
    &= \underline{p}^2 e^{(y_{q-p}-y_{p+\delta})} \\
    &= Q^2 \,.
\end{split}
\end{equation}
Also in the $q\bar{q}$ dipole we have often called the rapidity of the whole dipole $y$, labeling it by its softer component $y_{p+\delta}$.

\nocite{*}

\bibliography{refs}

\end{document}